\documentclass[aps,prb]{revtex4}
\usepackage{graphicx}
\usepackage{epsfig}
\usepackage{amsmath}
\usepackage{graphics}
\usepackage{epsfig}
\newcommand{\bra}[1]{\ensuremath{\langle{#1}|\,}}
\newcommand{\ket}[1]{\ensuremath{\,|{#1}\rangle}}

\begin{document}

\title{Super-fermion representation of quantum kinetic equations for the electron transport problem}
\author{Alan A. Dzhioev$^{1,2}$}
\author{D. S. Kosov$^{1} $}
\address{
$^1$Department of Physics and  Center for Nonlinear Phenomena and Complex Systems,
Universit\'e Libre de Bruxelles,
Campus Plaine, CP 231,
Blvd du Triomphe,
B-1050 Brussels,
Belgium }

\address{
${^2}$Bogoliubov Laboratory of Theoretical Physics, Joint Institute for Nuclear Research, RU-141980 Dubna, Russia }


\begin{abstract}
We discuss the use of super-fermion formalism to represent and solve quantum kinetic equations for the electron transport problem.
Starting with the Lindblad master equation for the molecule connected to two metal electrodes, we convert the problem of finding the nonequilibrium steady state  to the many-body problem with non-Hermitian  Liouvillian in super-Fock space. We  transform  the Liouvillian to the normal ordered form, introduce nonequilibrium quasiparticles
by  a set of canonical nonunitary transformations
and develop  general many-body theory for the electron transport through the interacting region. The approach is applied to the electron transport through a single level. We consider  a minimal basis hydrogen atom attached to two metal leads in Coulomb blockade regime (out of equilibrium Anderson model) within the nonequilibrium Hartree-Fock approximation as an example of the system with electron interaction. Our approach agrees with exact results given by the Landauer theory for the considered models.

\end{abstract}

\maketitle
\section{Introduction}

Recently, there have been a significant progress towards theoretical and experimental understanding of electron transport through single molecules.\cite{Agrait200381,nichols2010,nazarov-book,diventra-book}
The research has lead to  discovery of interesting physical transport phenomena in molecules such as Franck-Condon blockade,\cite{fcblockade05}  nonequilibrium Kondo effect,\cite{park02,natelson04,PhysRevLett.96.196601} negative differential resistance,\cite{Chen99a} vibronic effects and local heating,\cite{heating} as well as switching and hysteresis.\cite{galperin05}   Being a nonequilibrium system of
 strongly correlated electrons molecular junstions present many interesting and challenging  physical problems. Because of the very large current density
  (for example, if we consider a typical experimental current of a few $\mu A$ across the molecular junction of one atom wide, then the  current density would be much larger than in usual mesoscopic devices)\cite{diventra-book} and high inhomogeneity of electron density
 (such as, for example, cusps on nuclear positions or electron concentrations along  chemical bonds), the  electron-electron correlations play a pivotal role in electron transport through molecules.

 Much of the theoretical and computational studies of transport properties of molecular nanostructures are based on Landauer theory which
defines the conductance from the electron transmission probability.\cite{Landauer70} The transmission probability is usually   computed by
nonequilibrium Green's functions (NEGF).\cite{Caroli71} Most of the currently employed first principles electron transport  calculations
combine NEGF or scattering theory type approaches with ground state
density functional theory (DFT). \cite{Taylor01,Xue02,Brandbyge02,diventra:979,Fujimoto03, Ke04}
There are many diverse practical implementations of NEGF-DFT  and it  has been applied to various molecular junctions.\cite{Taylor01,Xue02,arnold:174101,Brandbyge02,thygesen:111,li:1347,smeagol, Fujimoto03,Calzolari2004,dunietz08}
The  achieving of the agreement between these  first principles  electron transport
 calculations and experiments has been elusive target for the last decade:
the theoretically predicted current   is systematically orders of magnitude too large. \cite{diventra:979, Evers04,  Xue02,li:035415,Burke-rev08,LiZ._jp065120t}
Since reliable and accurate experimental measurements of single molecule transport properties are now available,\cite{nichols2010,venkataraman:458} we cannot anymore blame  poorly controlled experiments for these discrepancies.  All practical applications
are limited to mean field,  and, in spite of some recent efforts,\cite{PhysRevB.77.115333,PhysRevB.80.115107,PhysRevB.79.155110,PhysRevB.80.165305,Baer03}   electron-electron interaction effects are difficult to include into practical calculations.
Therefore, we think,  it is very important at the current stage of research to move beyond currently employed NEGF-DFT scheme towards successful highly accurate methods for quantum chemical electronic structure calculations, such as, for example, coupled cluster theory\cite{cizek69,bartlett:291}
and configurational interaction.\cite{szabo-book} To succeed we need to develop a formalism which enables us to extend  ideas of these
sophisticated methods for equilibrium electronic structure calculations to nonequilibrium systems, namely, we ideally would like to reduce the problem of electron transport calculations to the eigenvalue problem with large, non-hermitian, but nevertheless finite matrix. This paper presents our first step in this direction.

The quantum kinetic equations provide a more suitable language for molecular type electronic structure calculations than NEGF or scattering theory based approaches. Recent progress in the development of kinetic equations for the electron transport presents a significant promises in this direction.\cite{gurvitz96,PhysRevB.78.235424,PhysRevB.74.235309,PhysRevB.80.045309,PhysRevB.71.205304,PhysRevB.72.195330,ovchinnikov:024707,subotnik:144105,mazziotti10}  Although to fully engage transport kinetic equations with the high level electronic structure calculations we need to transform them to the familiar language of second quantization, creation and annihilation operators, normal ordering, Fock space and vacuum. In this paper paper we develop a systematic scheme to convert and solve quantum kinetic equations in the language of the advanced quantum chemistry. Our derivation is based on the Lindblad master equation,\cite{Lindblad1976} but the approach can be readily extended to other types of kinetic equations.

We would like to distance the formalism presented in this paper from the recent work of one of the authors.\cite{kosov:171102} There was proposed a theoretical approach to the electron transport problem based on methods of equilibrium thermofield dynamics. The approach used the square root of the density matrix and  employs unitary canonical transformations to build nonequilibrium Fock space.
Because of that the main restriction of the method is the necessity to use  the full space  Liouvillian, which includes the bath continuum states, throughout the calculations.  Contrary, in the current  paper we develop a set of nonunitary canonical transformations which enables us to lift the above mentioned restriction.

The important remark on the notation used in the paper is the following: only creation/annihilation operators written with letter $a$ (such as for example $a_{k \alpha}$ and $a^{\dagger}_{k \alpha}$) are related to each other by the hermitian conjugation; all other creation  ($c^{\dagger}$, $b^{\dagger}$, $d^{\dagger}$ etc) and annihilation ($c$, $b$, $d$ etc) operators are "canonically conjugated" to each other, i.e. for example  $b^{\dagger}$ does not mean $(b)^{\dagger}$ although $\{ b,b^{\dagger}\}=1$.

The rest of the paper is organized as follows. In Section II, the aspects of the super-operator formalism relevant to the electron transport problem are described. We also demonstrate in this section how to solve the Lindblad master equation in the super-fermion space for a single impurity connected to one thermal bath. Section III presents the general formalism to work with the Lindblad  master equation in the super-fermion Fock space for the electron transport through the interacting region. In section IV, we apply the formalism to the electron transport through  single level and through molecule in  Coulomb blockade regime. In this section we also compare the results with the Landauer theory. Conclusions are given in Section V. We use natural units throughout the paper: $\hbar= k_B = |e| = 1$, where $-|e|$ is the electron charge.

\section{Liouville space and superoperator formalism}

In this section we describe the algebraic structure of the super-fermion space relevant to the electron transport calculations. We also establish the notation and terminology which are used throughout the paper. This section is partly based in the formalism of superoperators developed by   Schmutz  in
the context of real-time Green's functions method\cite{schmutz78}. Some aspects of the superoperator nonequilibrium Green's function theory with application to electron transport have been recently discussed by Harbola and Mukamel. \cite{Harbola2008} In contrast to Harbola and Mukamel's work,  the present  approach does not aim to  re-formulate Keldysh  NEGF in terms of super-operators and uses the  formalism to solve quantum kinetic equations for the electron transport problem.
We also use some ideas laid down by Prosen in his works on "third quantization" technique,\cite{prosen08,prosen2010}
where the similar method was used to solve master equation for non-interacting open Fermi systems with  quadratic Hamiltonians and  one dimensional spin chains.

A nonequilibrium quantum system can be described in terms of density matrix $\rho(t)$ which satisfies the Liouville equation:
\begin{equation}
i \dot{\rho}(t) = \left[H, \rho(t) \right].
\label{liouville}
\end{equation}
The average value of an operator at particular moment $t$ is given by
\begin{equation}
\langle A(t) \rangle  =   \mbox{Tr} \left( \rho(t) A \right) .
\label{neav}
\end{equation}
For many-particle quantum systems the density matrix $\rho(t)$ and the Hamiltonian are operators in the Fock space. The Fock space can be defined by some orthonormal complete set of basis vectors:
\begin{equation}
\sum_n |n \rangle \langle n| =I,\;  \;\;  \langle n| m \rangle = \delta_{nm}.
\label{fock}
\end{equation}
Let us introduce the additional Fock space which is identical copy of the initial Fock space
\begin{equation}
\sum_n |\widetilde{n} \rangle \langle \widetilde{n}| =\widetilde{I},\;  \;\;  \langle \widetilde{n}| \widetilde{m} \rangle = \delta_{nm}.
\label{tilde-fock}
\end{equation}
We denote all vectors and operators in this additional Fock space by "tilde".
The vectors $|n\rangle$ and $|\widetilde{n}\rangle $ span the so-called super-Fock space, which is a direct product of the original and the "tilde" Fock spaces. Operators in this super-Fock space will be called super-operators.
Let us also introduce  "left vacuum vector" (why it is appropriate to call it vacuum will be clear later)
\begin{equation}
|I \rangle = \sum_{n} |n \rangle \otimes  |\widetilde{n}\rangle,
\label{unit-vector}
\end{equation}
and  "nonequilibrium wavefunction":
\begin{equation}
|\rho(t)\rangle =\rho(t) | I \rangle
=\sum_{nm} \rho_{mn}(t)  |n\rangle \otimes  |\widetilde{m}\rangle,
\label{vacuum}
\end{equation}
where $\rho_{mn}(t) = \langle n | \rho(t) |m \rangle $. From ${\rm Tr}\,\rho(t)=1$ it follows that the left vacuum vector and the nonequilibrium wavefunction are ortonormal to each other, $\langle I |\rho(t)\rangle=1$. With the use of the  left vacuum  and the nonequilibrium wavefunction we can re-write the average (\ref{neav})
as  the following matrix element:
 \begin{equation}\label{expect}
 \langle A(t) \rangle = \langle I | A  |\rho(t)\rangle .
\end{equation}
The Liouville equation (\ref{liouville})
becomes equivalent to the time-dependent Schr\"odinger equation
in the super-Fock space
\begin{equation}
i\frac{d}{dt} |\rho(t)\rangle = {L} |\rho(t)\rangle,
\label{Schr_eq}
\end{equation}
where  ${ L} = H- \widetilde{H} $ is the Liouville superoperator (Liouvillian).
 The left vacuum is always  an eigenvector with  zero eigenvalue of the Liouvillian $\bra{I} L =0$, which also  automatically guarantees that $\langle I |\rho(t)\rangle=1$.

Let us consider a system which consists of  fermions distributed over $N$ levels. Let us take vector $|n\rangle$ and $|\widetilde{n}\rangle$  to be the particle number eigenstate
$|n \rangle = |n_1 n_2 ... n_N\rangle$ and $|\widetilde{n} \rangle = \widetilde{ |n_1 n_2 ... n_N\rangle}$
\begin{equation}
a^{\dagger}_i a_i |n_1 n_2 ... n_N \rangle = n_i |n_1 n_2 ... n_N \rangle.
\end{equation}
\begin{equation}
\widetilde{a}^{\dagger}_i \widetilde{a}_i \widetilde{|n_1 n_2 ... n_N \rangle} = n_i \widetilde{|n_1 n_2 ... n_N \rangle}.
\end{equation}
Here $a^{\dagger}_i$ ($a_i $) are fermion creation(annihilation) operators, which satisfy the standard anticommutation relations:
\begin{equation}
\{a_i, a_j^{\dagger}\} = \{\widetilde{a}_i, \widetilde{a}_j^{\dagger}\} =\delta_{ij}, \;\; \{a_i, a_j\} =\{a^{\dagger}_i, a^{\dagger}_j\} =
\{\widetilde{a}_i, \widetilde{a}_j\} =\{\widetilde{a}^{\dagger}_i, \widetilde{a}^{\dagger}_j\} = 0,
\end{equation}
and fermionic operators in initial and tilde Fock spaces anticommute with each other.
Then the left vacuum
can be written as
\begin{eqnarray}
|I \rangle =&&\sum_{n_1 n_2 ...n_N} |n_1 n_2 ... n_N \rangle \otimes \widetilde{ | n_1 n_2 ... n_N \rangle }
\nonumber
\\
=&&\exp(\sum_{i} a^{\dagger}_i  \widetilde{a}^{\dagger}_i )
 |0\rangle \otimes |\widetilde{0} \rangle,
\end{eqnarray}
where $ |0\rangle$  and $|\widetilde{0} \rangle$ are vacuums in ordinary and tilde Fock spaces (i.e. $a_i |0\rangle =0$ and $\widetilde{a}_i |\widetilde{0}\rangle =0$ for all $i$).
The vector $|n\rangle \otimes |\widetilde{n} \rangle $ can be always multiplied by some phase factor $\exp(i \alpha)$. We found that, when one works with the fermionic systems, the convenient choice of the phase  is the following
\begin{eqnarray}
&&|I \rangle =\sum_{n_1 n_2 ...n_N} (-i)^{n_1 + n_2 + ...+ n_N}  |n_1 n_2 ... n_N \rangle \otimes  \widetilde{|{n}_1 {n}_2 ... {n}_N \rangle}
\nonumber
\\
&&=\exp(-i \sum_{i} a^{\dagger}_i  \widetilde{a}^{\dagger}_i )
 |0\rangle \otimes |\widetilde{0} \rangle.
\end{eqnarray}
One can readily demonstrate by the straightforward algebraic manipulations that
\begin{equation}
a_j |I\rangle = -i \widetilde{a}^{\dagger}_j |I\rangle, \;\;\;\;\; a^{\dagger}_j |I\rangle = -i \widetilde{a}_j |I\rangle.
\label{tilde}
\end{equation}
Following the terminology of the thermofield dynamics \cite{matsumoto85,umezawa} we will call the above relations as the "tilde conjugation rules".
The above relation which transform the original operators to the tilde operators is one of the most important relations and it is used very often in our derivations in this paper. In particular,  it follows from~\eqref{tilde} that the left vacuum  in the super-Fock space, $\bra{I}$, is the vacuum for $a^\dag_j-i\widetilde{a}_j$  and $\widetilde a^\dag_j+i a_j$ operators. Since also $\bra{I} L =0$,  it is appropriate to call $\langle I | $ left vacuum vector.

We will focus  in this paper on the  Lindblad quantum master equation.\cite{Lindblad1976}  It can be written in the following general form
\begin{eqnarray}
\label{L_eq}
  \frac{d}{dt}\rho(t) =-i[H,\rho(t)]+\sum_\mu(2L_\mu\rho(t) L^\dag_\mu
   -\{ L^\dag_\mu L_\mu,\rho(t)\}),
\end{eqnarray}
where $H$ is the Hamiltonian of the system, $L_{\mu}$ is a set of generally non-Hermitian Lindblad operators that
represent the influence of the environment on the system, and  $\{ L^\dag_\mu L_\mu,\rho(t)\}$ means the anticommutator $L^\dag_\mu L_\mu \rho(t) + \rho(t) L^\dag_\mu L_\mu$. The Lindblad master equation is the most general master equation which can be derived under the requirements that all probabilities are real and nonnegative, $\rho(t)$ is always normalized, and $\rho(t)$ can be obtained from $\rho(0)$ by linear map.\cite{Petruccione,Lindblad1976}
The last term  in the Lindblad equation is so-called  dissipator. It contains the term $L_\mu\rho L^\dag_\mu$ which creates quantum jumps between the states of the system and the  term $\{ L^\dag_\mu L_\mu,\rho\}$ balances the quantum fluctuations from the quantum jumps.

 Let us understand how we can re-write and solve the Lindblad master equation in super-fermion representation. We consider the following simple but  nonetheless  important example: A single level connected to  thermal bath with temperature $T$  and chemical potential~$\mu$.
The Hamiltonian is
 \begin{equation}
    H= \varepsilon a^\dag a
 \end{equation}
and we take Lindblad operators in the following form\cite{prosen08}
 \begin{equation}
    L_1=\sqrt{\Gamma_1}a,~~~L_2=\sqrt{\Gamma_2}a^\dag.
 \end{equation}
The Lindblad equation becomes
\begin{eqnarray}
\label{linb}
  &&  \frac{d}{dt}\rho(t)=-i \varepsilon (a^\dag a\rho-\rho a^\dag a)+\Gamma_1(2a\rho a^\dag - a^\dag a\rho-\rho a^\dag a)
    \nonumber
    \\
&& +    \Gamma_2(2a^\dag\rho a - a a^\dag\rho-\rho a a^\dag)
\end{eqnarray}
If we act by  this equation on the left vacuum $|I\rangle$, use the tilde conjugation rules (\ref{tilde}) and consider that the density matrix
$\rho= \rho(a^{\dagger}, a) $ is the operator in original Fock space therefore it commutes with all tilde operators,
we obtain the  time-dependent Schr\"odinger equation (\ref{Schr_eq}) for the nonequilibrium wavefunction $\ket{\rho(t)}$
 with the Liouvillian ${L}$ given by
 \begin{eqnarray}
 \label{therm_H}
L&&=\varepsilon (a^\dag a-\widetilde a^\dag \widetilde
a)-i(\Gamma_1-\Gamma_2)(a^\dag a+\widetilde a^\dag \widetilde
a)
\nonumber
\\
&&-2(\Gamma_1 \widetilde a a+\Gamma_2 \widetilde a^\dag
a^{\dagger})-2i\Gamma_2.
 \end{eqnarray}

Within nonequilibrium thermofield dynamics a similar expression for Liouvillian (in nonequilibrium thermofield dynamics it is called
Tildian) in super-Fock space was obtained axiomatically (see for example review~\cite{Kobryn2003} ).

Let us compute the average number of electrons on level $\varepsilon$ at time $t$:
 \begin{equation}\label{distr}
    n(t)=\langle I|a^\dag a|\rho(t)\rangle.
 \end{equation}
Differentiating this equation with respect to time we get
\begin{equation}
   i \frac{d}{d t} n(t)=\langle I |a^\dag a L |\rho(t)\rangle.
    \label{dndt}
\end{equation}
Since $\bra{I}L=0$ the right side  of Eq.~\eqref{dndt} can be written as the commutator $\langle I |[a^\dag a ,
    L ]|\rho(t)\rangle$.
    Direct calculation of this commutator with the Liouvillian (\ref{therm_H}) results into the following equation for the time-evolution of the occupation number
\begin{equation}
    \frac{d}{d t} n(t)=-2(\Gamma_1+\Gamma_2)\{
    n(t)-\frac{\Gamma_2}{\Gamma_1+\Gamma_2}\},
\end{equation}
which has the solution
\begin{equation}\label{nt}
    n(t)=(n_0-n_\infty){\rm e}^{-2(\Gamma_1+\Gamma_2)t}+n_\infty.
\end{equation}
Here $n_0$ is the initial occupation number at  $t=0$,
$n_\infty=\frac{\Gamma_2}{\Gamma_1+\Gamma_2}$ -- asymptotic occupation number at  $t=\infty$.  Assuming
$|\rho_\infty\rangle$ corresponds to the equilibrium density matrix in the grand canonical ensemble, we get
\begin{equation}
\langle I | a^\dag a| \rho_\infty \rangle= f,
\end{equation}
where $f=[1+\exp[(\varepsilon-\mu)/T]^{-1}$ is the  Fermi-Dirac distribution function.  So
the choice  of the rates $\Gamma_{1}$ and  $\Gamma_{2} $ in the Lindblad operators  in the  form  $\Gamma_{1}=\gamma (1-f)$, $\Gamma_{2}=\gamma f$ leads to asymptotic equilibrium state in the grand canonical ensemble with temperature $T$ and chemical potential $\mu$.  The parameter $\gamma=\Gamma_{1}+\Gamma_{2}$ is
the relaxation rate to thermal equilibrium. The calculations along the same lines  for the Lindblad master equation were also performed by
Prosen  within his "third quantization" method.\cite{prosen08}

Let us summarize all our observation as a set of  practical  rules for the operations in the super-Fock space:
\begin{enumerate}
\item
The  left vacuum vector $| I \rangle $ and the nonequilibrium wavefunction $ |\rho(t)\rangle =\rho(t) | I \rangle $  are invariant under the tilde conjugation
$\widetilde{|\rho(t) \rangle }  = |\rho(t) \rangle $,  $\widetilde{| I \rangle }  = | I \rangle $, and $\bra{I}\rho(t)\rangle=1$.

\item Tilde conjugation rules: The main rule is
$
a_j |I\rangle = -i \widetilde{a}^{\dagger} |I\rangle, \;\;\; a^{\dagger}_j |I\rangle = -i \widetilde{a} |I\rangle
$
and as consequence the double tilde conjugation does not change the operator $\widetilde{\widetilde{A}}  =A$
and
$
\widetilde{(c_1 A + c_2 B)} = c_1^* \widetilde{A} + c^*_2 \widetilde{B}
$.

\item Evolution of the system is described by the time-dependent Schr\"odinger equation
$
i\frac{d}{dt} |\rho(t)\rangle = L |\rho(t)\rangle,
$
where the Liouvillian is obtained from the  corresponding master equation for the density matrix with help of  the tilde conjugation rules.
The nonequilibrium average is given by $\langle A(t) \rangle = \langle I | A |\rho(t) \rangle $ and $ \langle I | L =0$.

\end{enumerate}

\begin{figure}
 \begin{centering}
\includegraphics[width=10cm]{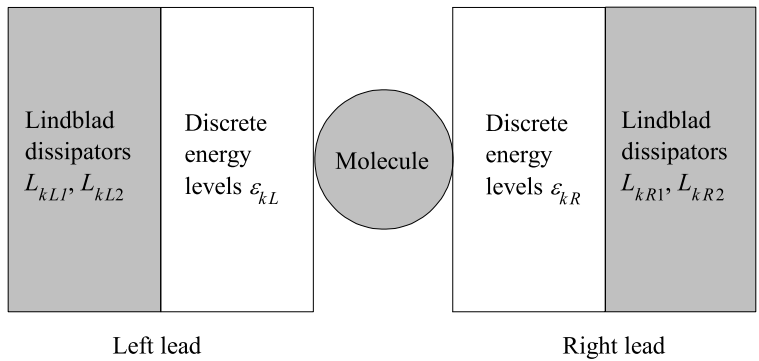}
\caption{ Schematic illustration of the partitioning of the system. The left and right leads are assumed to be described by the non-interacting electrons, whereas electron-electron  interaction is present
in the molecule.
Each lead is separated into two parts: The finite parts connected directly to the molecule  and the macroscopically large parts (grey area) are represented by the Lindblad dissipators. }
 \label{figure0}
 \end{centering}
\end{figure}

\section{Quantum master equations for electron transport through the interacting region in super-fermion representation}\label{section3}

Let us now apply the formalism of superoperators discussed in the previous section to the problem of electron transport through the
molecular junction.
If we place a  molecule into the
contact with two electron  reservoirs (usually metal leads) with different chemical potentials, the electric current starts to flow through it.
Then, if wait for some time, which is much longer than  typical electronic relaxation time of the molecule,
the system  will reach a  nonequilibrium time-independent steady state. In this section, we will discuss some general theoretical ideas of electron transport through the molecule in nonequilibrium steady state  regime based on the super-fermion representation of Lindblad master equation.

We begin with the  tunneling Hamiltonian:
\begin{equation}
H= H_S(a^{\dagger}_s, a_s) + \sum_{k, \alpha= L,R}  \varepsilon_{k \alpha} a^{\dagger}_{k \alpha} a_{k \alpha} -
\sum_{s, k, \alpha= L,R } (t_{s k \alpha } a^{\dagger}_{k \alpha} a_s + h.c.),
\label{H_tot}
\end{equation}
where $a^{\dagger}_{k\alpha}$ ($a_{k\alpha}$) creates (destroys) an electron in the state $k$ of either the left ($\alpha=L$) or  the right ($\alpha=R$) lead,
$a^{\dagger}_s$ and  $a_s$ are creation annihilation operators in the molecule, and $t_{sk \alpha}$ are the coupling parameters between the system and the leads.
In what follows we assume $t_{sk \alpha}=t^*_{sk \alpha}$. Here $H_S(a^{\dagger}_s, a_s) $  represents the molecule  and contains two-particle electron-electron correlations and, if necessary, electron-vibration coupling.
The electron creation and annihilation operators $a_s^\dag~(a_s)$ and $a^\dag_{k\alpha}~(a_{k\alpha})$
satisfy  standard  anticommutation relations.  We approximate each lead by
$N$ {\it discrete} single particle levels~$k$ coupled to the Lindblad dissipators (see Fig.~{\ref{figure0}).
We assume that if  two leads were disconnected from each other, they would be in thermodynamic equilibrium with temperatures $T_{L}$ and $T_{R}$ and chemical
potentials~$\mu_{L}$ and $\mu_{R}$. As  follows from the previous section, it can be accomplished by the use of the following set of $4N$ Lindblad operators:
 \begin{equation}\label{L_op}
 L_{k\alpha1}=\sqrt{\Gamma_{k\alpha1}}a_{k\alpha},~~~L_{k\alpha2}=\sqrt{\Gamma_{k\alpha2}}a^\dag_{k\alpha}
\end{equation}
with $\Gamma_{k\alpha1}=\gamma_{k\alpha}(1- f_{k\alpha})$, $\Gamma_{k\alpha2}=\gamma_{k\alpha} f_{k\alpha}$ and
 $f_{k\alpha}=[1+\exp[(\varepsilon_{k\alpha}-\mu_\alpha)/T_{\alpha}]^{-1}$. In the present model, the bias voltage changes only at the interface between the interacting region and the leads.
 We also think that a  particular choice of the dissipators is not an issue here as long as  we can place the boundary between the part of the lead which is represented by the Lindblad dissipator and part of the lead which is given by the discrete single particle levels (Fig.~{\ref{figure0}) deep enough  inside the metal. This partitioning is justified in realistic systems since the screening length in the metallic leads is very short.

 The Lindblad master equation  is
 \begin{eqnarray}
\label{lindblad}
  \frac{d\rho(t)}{dt} =-i[H,\rho(t)]+\sum_{k\alpha} \sum_{\mu=1,2} (2L_{k \alpha \mu } \rho(t) L^\dag_{k \alpha \mu}
   -\{ L^\dag_{k \alpha \mu} L_{k \alpha \mu},\rho(t)\}).
\end{eqnarray}
In principle, this Lindblad master equation (\ref{lindblad}) for the electron transport can be obtained directly from the general full space tunneling Hamiltonian under some quite general assumptions by projecting out parts of the leads.\cite{gurvitz96,PhysRevB.74.235309}

Now we would like to convert this Lindblad master equation to the super-Fock space. Likewise to the single-level in equilibrium considered in the previous section,
we act by  the Lindblad equation on the left vacuum vector $|I\rangle$, employ  tilde conjugation rules  and use the fact that the density matrix
$\rho= \rho(a^{\dagger}, a) $ is an operator in the original Fock space therefore it commutes with all tilde operators. Then
the Lindblad master equation becomes the time-dependent Schr\"odinger equation in the super-Fock space (\ref{Schr_eq}) with the following non-Hermitian Liouvillian
\begin{align}\label{H_lind}
 L& =  L_B +  L_S + L_T.
\end{align}
Here
\begin{equation}
L_S = H_S-\widetilde H_S
\end{equation}
is the Liouvillian for the molecule
and
\begin{equation}
L_T =  -\sum_{s k \alpha}t_{s k \alpha } ( a^{\dagger}_{k \alpha} a_s -\widetilde{a}^{\dagger}_{k \alpha} \widetilde{a}_s+ h.c.)
\end{equation}
is the Liouvillian, which describes the coupling between the molecule and the leads. They are both Hermitian.
The Liouvillian for the leads
\begin{equation}
L_B = \sum_{k \alpha} \varepsilon_{k \alpha} (a^{\dagger}_{k \alpha} a_{k \alpha} - \widetilde{a}^{\dagger}_{k \alpha} \widetilde{a}_{k \alpha}) -i
 \sum_{k\alpha}\Pi_{k\alpha}
\end{equation}
includes the non-Hermitian part of the Liouville operator which is responsible for the dissipation in the system:
\begin{align}
\Pi_{k\alpha} =(\Gamma_{k\alpha1}-\Gamma_{k\alpha2})(a^\dag_{k\alpha}a_{k\alpha}+
  \widetilde a^\dag_{k\alpha}\widetilde a_{k\alpha})-
  2i(\Gamma_{k\alpha1}\widetilde a_{k\alpha}a_{k\alpha}+\Gamma_{k\alpha2}\widetilde a^\dag_{k\alpha}a^\dag_{k\alpha})+2\Gamma_{k\alpha2}.
  \label{Pi}
  \end{align}
 It comprises the terms  such as $a_{k \alpha} \widetilde{a}_{k \alpha} $ and  $a^{\dagger}_{k \alpha} \widetilde{a}^{\dagger}_{k \alpha}$ which usually appear in the theory of superfluidity.\cite{bogoliubov58-hfb} Due to these terms  the structure of the nonequilibrium steady state wavefunction for the electron transport problem will have mathematical similarities with the vacuum in the Hartree-Fock-Bogoliubov theory.\cite{bogoliubov58-hfb} This terms originate from the part of the Lindblad dissipator which is responsible for the "quantum jumps" between single-particle energy levels $\varepsilon_{k\alpha}$ in the leads.

We are interested in nonequilibrium steady state  situation, where the density matrix $\rho(t)$  or nonequilibrium wavefunction $|\rho(t)\rangle $ does not depend on time. Therefore, the electron transport problem is reduced to the problem of  finding the eigenvector
with zero eigenvalue of complex, non-Hermitian, finite-dimensional Liouville operator
\begin{equation}
L |\rho_\infty\rangle =0.
\end{equation}
Here $|\rho_\infty\rangle$ is nonequilibrium  wavefunction for the nonequilibrium steady state.
Subscript $"\infty"$ in $|\rho_\infty\rangle$ serves to emphasize that the nonequilibrium  steady state can be also considered as asymptotic ($t \rightarrow \infty $) state of the system.

Suppose we have diagonalized the Liouvillian~\eqref{H_lind} in some $\xi$-modes exactly or within some approximation
\begin{equation}
L=\sum_n (\Xi_n \xi^\dag_n\xi_n - \Xi^*_n \widetilde\xi^\dag_n\widetilde \xi_n ) .
\end{equation}
Suppose also that $\bra{I}\xi^\dag_n=\bra{I}\widetilde\xi^\dag_n=0$. Since $\bra{I}{L}=0$, then after  diagonalization ${L}$ does not contain "c"-number terms. As a result a nonequilibrium steady
 state can be determined as a vacuum of $\xi$-modes. The creation and annihilation operators of  $\xi$-modes  will not be Hermitian adjoint to each other, since the left and right vacuums are different.  If we recall that
 the basic idea of standard (equilibrium) quasiparticles is to represent true ground state of interacting many particle systems as  a vacuum with respect to some quasiparticle annihilation  operators,\cite{landau56,bogoliubov58}
 then  these $\xi$-modes can be regarded as nonequilibrium quasiparticles.
The nonequilibrium quasiparticles include
nonequilibrium, correlations and dissipation  into their structure.
 This nonequilibrium quasiparticle description is, in principle, exact if one defines  the quasiparticles in terms of the exact eigenstates of the many-particle Liouvillian. Although in practice it can
be done  approximately by establishing  the relation between quasiparticles and "bare" particles of the system, for example, by  canonical  transformations. Below, we demonstrate how it can be done.

 Let us begin diagonalization of the Liouvillian (\ref{H_lind}). First, we separate the part of the Liouvillian which contains only operators from the leads $L_B$.
  It is not Hermitian due to the presence of the dissipators.
 Direct calculations
 show that the following canonical (but not unitary) transformation
\begin{align}\label{transf2}
&  a_{k\alpha}=b_{k\alpha}-if_{k\alpha}\widetilde b^\dag_{k\alpha},~~~ \widetilde a_{k\alpha}= \widetilde b_{k\alpha} + if_{k\alpha}  b^\dag_{k\alpha},
 \notag  \\
&a^\dag_{k\alpha}=(1-f_{k\alpha})b^\dag_{k\alpha}+i\widetilde b_{k\alpha},~~ ~\widetilde a^\dag_{k\alpha}=  (1 - f_{k\alpha}) \widetilde b^\dag_{k\alpha}  - i b_{k\alpha} ,
\end{align}
 with  $f_{k\alpha}=\Gamma_{k\alpha 2}/(\Gamma_{k\alpha 1}+\Gamma_{k\alpha 2})= [1+\exp[(\varepsilon_{k\alpha}-\mu_\alpha)/T_{\alpha}]^{-1}$, brings  $L_B $ to   the diagonal form
\begin{equation}
L_B=\sum_{k\alpha}\left\{
 E_{k\alpha} b^\dag_{k\alpha}b_{k\alpha}-E^*_{k\alpha}\widetilde b^\dag_{k\alpha}\widetilde b_{k\alpha}\right\} .
 \label{hbath}
\end{equation}
Here $E_{k\alpha}=\varepsilon_{k\alpha}-i(\Gamma_{k\alpha1}+\Gamma_{k\alpha2})=\varepsilon_{k\alpha}-i\gamma_{k\alpha}$ and $E^*_{k\alpha} = (E_{k\alpha})^* $ is complex conjugated energy. Nonunitary canonical transformations mean that these transformations, although being nonunitary, preserve the anticommutation relation between the fermion creation/annihilation operators.
Since,  $b^\dag_{k\alpha}=a^\dag_{k\alpha}-i\widetilde a_{k\alpha}$, the vector $\bra{I}$  is automatically the vacuum
for $b^\dag_{k\alpha}$ and $\widetilde b^\dag_{k\alpha}$ operators.
 We would like to repeat the important remark on the notation we use: only creation/annihilation operators written with letter $a$ (such as, for example, $a_{k \alpha}$ and $a^{\dagger}_{k \alpha}$) are related to each other by Hermitian conjugation; all other creation ($b^{\dagger}$, $c^{\dagger}$,   etc.) and annihilation operators ($b$, $c$,  etc.) are "canonically conjugated" to each other , i.e. for example  $b^{\dagger}$ does not mean $(b)^{\dagger}$ although $\{ b,b^{\dagger}\}=1$.

Having reduced the reservoir part to the ideal gas of non-Hermitian fermions, let us begin to work with Liouvillian of the interacting region.
We take the Hamiltonian  for the molecule in the following general form
\begin{equation}\label{H_gen}
  H_S=\sum_{s_1s_2} K_{s_1 s_2} a^\dag_{s_1} a_{s_2}+\frac14 \sum_{s_1 s_2 s_3 s_4}V_{s_1 s_2 s_3 s_4}a^\dag_{s_1} a^\dag_{s_2} a_{s_4} a_{s_3},
\end{equation}
where $K_{s_1 s_2} $ is  single-particle matrix element which contains  kinetic energy, electron-nuclei attraction and interaction with external fields such as, for example, gate voltage,  $ V_{s_1 s_2 s_3 s_4} $ is  antisymmetrized matrix elements of Coulomb electron-electron interactions.  Using the Wick theorem we perform the normal ordering
of the Liouvillian for the interacting region:
\begin{equation}
 L_S = H_S-\widetilde H_S=L^{(0)}_{S}+L'_S,
\end{equation}
where
 $ L^{(0)}_S$ is the quadratic part
 \begin{equation}\label{sp_HF}
  L^{(0)}_{S}=\sum_{s_1 s_2}\bigl(K_{s_1 s_2}+\sum_{s_3 s_4} n_{s_4 s_3}V_{s_1 s_3 s_2 s_4}\bigr):a^\dag_{s_1} a_{s_2} :- \rm{t.c.},
 \end{equation}
 and $L'_S$
contains part of the electron-electron interaction, which is irreducible to the quadratic form :
\begin{equation}
  L'_{S}=\frac14\sum_{s_1 s_2 s_3 s_4}\left( V_{s_1 s_2 s_3 s_4}:a^\dag_{s_1} a^\dag_{s_2} a_{s_4} a_{s_3}: -  \rm{t.c.}  \right).
\end{equation}
 The steady state single-particle density matrix is $n_{s_2 s_1}=\bra{I} a^\dag_{s_1} a_{s_2} $\ket{\rho^{(0)}_\infty}, and the notation (t.c.) means the tilde conjugation (i.e. $a_s \to \widetilde a_s$, $n_{s_2s_1}\to n^*_{s_2s_1}$, etc.).
The normal ordering is asymmetric: it is performed with respect to  the left vacuum $\langle I|$ from the left and nonequilibrium vacuum $ \ket{\rho^{(0)}_\infty}$, which is the solution of the following eigenvalue problem
\begin{equation}
L^{(0)} \ket{\rho^{(0)}_\infty} = 0,
\label{hf}
\end{equation}
from the right. Here
\begin{equation}
L^{(0)}=L^{(0)}_S+L_T+ L_B
\end{equation}
is the quadratic part of the  Liouvillian. We would like to comment on the applicability of the Wick theorem to our case, when the left vacuum state is not the same as the right vacuum.  As it will be demonstrated below, we can always write operators $a_s, a^{\dagger}_s$ as a linear combinations of some fermionic creation and annihilation operators (let us call them $c_n$ ) in such a way that $ c_n \ket{\rho^{(0)}_\infty}  = \widetilde{c}_n \ket{\rho^{(0)}_\infty}=0$, and  $  \langle I | c^{\dagger}_n  = \langle I | \widetilde{c}^{\dagger}_n=0 $. In this case,
 the Wick  theorem is applicable\cite{bogoliubov-shirkov} and the Liouvillian can be brought to the normal form with respect to the "left vacuum" $\bra{I}$  from the left and  the vacuum for the quadratic part of the Liouvillian $ \ket{\rho^{(0)}_\infty}$  (\ref{hf}) from the right.

Since $L^{(0)}_S$ is quadratic and Hermitian, it can be diagonalized  exactly by unitary transformation $\mathbf{D}$ of  creation and annihilation operators
\begin{equation}\label{transfHF}
a_{s} = \sum_{s'} D_{s s'} \alpha_{s'}
\end{equation}
Note, that $\mathbf{D}$ does not mix nontilde and tilde operators. To diagonalize~\eqref{sp_HF}, $L^{(0)}_S=\sum_s \varepsilon_s (\alpha^\dag_s \alpha_s - \widetilde\alpha^\dag_s \widetilde \alpha_s)$, the matrix  $\mathbf{D}$ should satisfy the following equation
\begin{equation}
  \sum_{s_2}\bigl(K_{s_1 s_2}+\sum_{s_3 s_4} n_{s_4 s_3}V_{s_1 s_3 s_2 s_4}\bigr)D_{s_2s}=\varepsilon_s D_{s_1s}.
\end{equation}
This is the nonlinear eigenproblem, since  $n_{s_4s_3}$ depends on $\mathbf{D}$ and $\varepsilon_s$.  Next,
we introduce  system operators ${b^\dag_s=\alpha^\dag_s - i\widetilde\alpha_s}$, ${\widetilde b^\dag_s=\widetilde\alpha^\dag_s + i\alpha_s}$  which annihilate left vacuum $\bra{I}$ due to tilde conjugation rules. The canonically conjugate annihilation operators are
 $b_s=\alpha_s,~~\widetilde b_s=\widetilde a_s$.  Then the  Liouvillian takes the form
\begin{equation}\label{Liov}
L= \sum_s \varepsilon_s (b^\dag_s b_s - \widetilde b^\dag_s \widetilde b_s) + L_B +  L_T + L'_S=L^{(0)}+L'_S,
\end{equation}
where the tunneling interaction is expressed in terms of  new creation and annihilation operators $b_{k \alpha }$, $b_s$
\begin{equation}
  L_T= -\sum_{s k\alpha  } T_{s k\alpha  }\left\{ (b^\dag_{k \alpha }b_s+b^\dag_s  b_{k \alpha }) -
  (\widetilde b^\dag_{k \alpha }\widetilde b_s+\widetilde b^\dag_s  \widetilde b_{k \alpha }) +
     i f_{k \alpha } (\widetilde b^\dag_{k \alpha}b^\dag_s  +  b^\dag_{k\alpha }\widetilde b^\dag_s) \right\},
\end{equation}
and $T_{s k \alpha }=\sum_{s'} t_{s' k\alpha } D_{ss'}$ are renormalized matrix elements of the tunneling interaction. After the transformation the tunneling part of  the Liouvillian depends  on temperatures and chemical potentials of the leads
through the Fermi-Dirac occupation numbers $f_{k \alpha} $.

Next we  diagonalize the quadratic part of~\eqref{Liov}, $L^{(0)}$, exactly,
\begin{equation}\label{L_op2}
L^{(0)}=\sum_{n} ( \Omega_n c^\dag_n c_n - \Omega^*_n \widetilde c^\dag_n \widetilde c_n).
\end{equation}
It results to the nonequilibrium Hartree-Fock theory for the electron transport problem. Creation and annihilation operators
 $c^\dag_n$, $c_n$  can be regarded as nonequilibrium quasiparticles with complex spectrum~$\Omega_n$ (the spectrum of tilde conjugated
nonequilibrium  quasiparticles is given by $-\Omega^*_n$). These operators obey the fermionic anticommutation relations, although   $c^\dag_n$ and $c_n$
 ($\widetilde c^\dag_n$ and $\widetilde c_n$) are  not Hermitian conjugated to each other.

To find the internal structure and energy spectrum of nonequilibrium quasiparticles we use the equation-of-motion method: If the Liouvillian is diagonal (\ref{L_op2}),
then  creation and annihilation operators $c_n$ and $c^{\dagger}_n$ must satisfy the following equations of motion
\begin{equation}\label{eq1}
[c^\dag_n,L^{(0)}]=-\Omega_n c^\dag_n,
\end{equation}
\begin{equation}\label{eq2}
[c_n,L^{(0)}]=\Omega_n c_n.
\end{equation}
Equations for $\widetilde c_n,~\widetilde c^\dag_n$ are obtained from~(\ref{eq1},\ref{eq2}) by  the tilde conjugation rule. We want to emphasize here
that, since $c_n\ne (c^\dag_n)^\dag $ and $\Omega_n$ is complex,  equations (\ref{eq1},\ref{eq2}) can not be
obtained from each other by the Hermitian conjugation.
To avoid unnecessary complication, we do not give a general solution of~Eqs.(\ref{eq1}, \ref{eq2}), we would  rather solve these equations for some particular
examples in the next section. Here, we only note that operators which diagonalize the quadratic part of~\eqref{Liov} have the following form
\begin{align}\label{c_gen}
  c^\dag_n&=\sum_s \psi_{n,\,s}b^{\dagger}_s+\sum_{k\alpha}\psi_{n,\,k\alpha}  b^\dag_{k\alpha},
   \notag \\
  c_n&=\sum_s (\psi_{n,\,s}b_s +  i \varphi_{n,\,s} \widetilde b^\dag_s)+\sum_{k\alpha}(\psi_{n,k\alpha}  b_{k\alpha} +i \varphi_{n,k\alpha} \widetilde  b^\dag_{k\alpha}),
\end{align}
and $\widetilde c^\dag_n$ and $\widetilde c_n$ are obtained from~\eqref{c_gen} by the  tilde conjugation rule.
Note that since the left vacuum $\bra{I}$ is the vacuum
for $b^\dag,~\widetilde b^\dag$ operators,  nonequilibrium quasiparticle creation operators $c^\dag_n,~\widetilde c^\dag_n$ are linear combinations of creation operators only.
With help of the anticommutation relations between nonequilibrium quasiparticle creation and annihilation operators, $\{c_n, c^\dag_{n'}\}=\delta_{nn'}$, $ \{ c_n, \widetilde c_{n'} \}=0$, we find that
the amplitudes $\psi$, ${\varphi}$ satisfy the following
orthogonality conditions:
\begin{align}\label{orthogonality}
  &\sum_s \psi_{n,\,s}\psi_{{n'},\,s}+\sum_{k\alpha} \psi_{n,\,k\alpha}\psi_{{n'},\,k\alpha}=\delta_{nn'},
     \notag \\
 & \sum_s (  \psi_{n,\,s}\varphi^*_{{n'},\,s}-\varphi_{n,\,s}\psi^*_{{n'},\,s})
      +\sum_{k\alpha}(  \psi_{n,\,k\alpha}\varphi^*_{{n'},\,k\alpha}-\varphi_{n,\,k\alpha}\psi^*_{{n'},\,k\alpha})=0.
\end{align}

By diagonalizing the Liouvillian we simultaneously find the nonequilibrium steady state $L^{(0)} |\rho_\infty^{(0)} \rangle =0$ as a (right) vacuum for operators $c_n$ and $\widetilde{c}_n$.
Using the transformation inverse to~\eqref{c_gen} we can express any  operator in terms of nonequilibrium quasiparticle operators. Then  all physical quantities are calculated
as an expectation value with respect to $\bra{I}$ and $|\rho_\infty^{(0)} \rangle$ vacuum states (see Eq.~\eqref{expect}) in the nonequilibrium Hartree-Fock approximation.
For the steady state current from the lead $\alpha=L,R$
\begin{equation}
  J^{(0)}_\alpha=-\frac{d}{dt} \sum_{k}  \bra{I} a^\dag_{k\alpha} a_{k\alpha} |\rho_\infty^{(0)} \rangle =
  -i \sum _{ks} t_{sk\alpha}\bra{I} ( a^\dag_{k\alpha} a_s -  a^\dag_s a_{k\alpha}) |\rho_\infty^{(0)} \rangle
\end{equation}
we derive, after some algebra, that
\begin{equation}\label{current_HF}
  J^{(0)}_\alpha= -i\sum _{ksn} T_{sk\alpha}(\psi^*_{n,k\alpha}\varphi^*_{n,s} -\psi^*_{n,s}\varphi^*_{n,k\alpha})=
  - 2 \mathrm{Im}  \sum _{ksn} T_{sk\alpha}\psi_{n,k\alpha}\varphi_{n,s},
\end{equation}
where the last equality follows from
\begin{equation}
  [b_s, \widetilde b_{k\alpha}]=\sum_n(\psi_{n,s}\varphi_{n,k\alpha}-\psi^*_{n,k\alpha}\varphi^*_{n,s})=0.
\end{equation}
For the steady state single-particle density matrix we have
\begin{equation}
n_{s_2s_1}=\sum_{s'_1,s'_2,n} D_{s_1s'_1}D_{s_2s'_2}\psi_{n,s'_1}\varphi_{n,s'_2}.
\end{equation}

Let us summarize the main results of this section. We have shown that  the quadratic part of the  Liouvillian~\eqref{H_lind}
can be diagonalized by  three  canonical transformations.
The first transformation  diagonalizes the
reservoir part of the Liouvillian including dissipator, the second is performed on the electrons in the interacting region and  diagonalizes the Hartree-Fock part of $L_S$, and the last transformation mixes  operators from the interacting region and the leads and diagonalizes the entire quadratic part of the Liouvillian. Two of these three transformations are not unitary,  but all of them are canonical, which means that the
anticommutation relations between particle creation and annihilation operators are preserved.  The remaining part $L'_{S}$
can be taken into account via standard perturbation theory or nonperturbatively by, for example, coupled cluster method\cite{bartlett:291} or  configuration interaction theory.\cite{szabo-book}
In fact, any other method for the correlated electronic structure calculations can be extended to nonequilibrium within our approach, although  special care should be taken because one has to work with the nonunitary representation of the creation and annihilation operators.

\section{Analytical and numerical examples}

\subsection{Transport through a single-level molecule}

To illustrate the theory
we consider  a single-level molecule  connected to two leads held at different chemical potentials. The molecule is described by the Hamiltonian
\begin{equation}
H_S=\varepsilon a^{\dagger} a.
\end{equation}
 In this section we obtain  nonequilibrium steady state wavefunction $\ket{\rho_\infty}$ for this model and compare the results with the Landauer theory. To simplify the notation  we assume throughout the calculations that the matrix elements of the tunneling interaction are real and do not depend on the leads energy levels, i.e., $t_{k\alpha}= t^*_{k\alpha} =t$. So the Liouvillian becomes:
\begin{equation}\label{liouv_single}
L= \varepsilon(a^\dag a-\widetilde a ^\dag \widetilde a) -
t \sum_{k \alpha} (a^{\dagger}_{k \alpha} a - \widetilde{a}^{\dagger}_{k \alpha} \widetilde{a}+ h.c.)
+ \sum_{k \alpha} \varepsilon_{k\alpha } (a^{\dagger}_{k \alpha} a_{k \alpha} - \widetilde{a}^{\dagger}_{k \alpha} \widetilde{a}_{k \alpha}) -i
 \sum_{k\alpha}\Pi_{k\alpha},
\end{equation}
where $\Pi_{k\alpha}$ is the dissipator for the lead $\alpha=L,R$ taken in the standard form (\ref{Pi}).
Our goal now is the  exact diagonalization of the Liouvillian (\ref{liouv_single}) in terms of nonequilibrium, non-Hermitian quasiparticles:
\begin{equation}
L=\sum_{n=1}^{2N+1} \left( \Omega_n  c_n^{\dagger} c_n -   \Omega^*_n  \widetilde{c}_n^{\dagger} \widetilde{c}_n \right).
\label{liouv-imp-d}
\end{equation}
To do this, we first perform the transformations (\ref{transf2}) over the leads operators and $b^\dag=a^\dag - i\widetilde a,~b=a$ for the molecular level. Then the equations of motion~\eqref{eq1}  give  the following structure for the   nonequilibrium quasiparticle creation and annihilation operators:
\begin{equation}\label{a_n2}
  c^\dag_n=\psi_{n}  b^{\dagger}+\sum_{k\alpha}\psi_{n,k\alpha}  b^\dag_{k\alpha},
\end{equation}
\begin{align}
\label{a_n22}
c_n&=\psi_{n} b +  i  \varphi_{n} \widetilde  b^\dag+\sum_{k\alpha}(\psi_{n,k\alpha}  b_{k\alpha} +i \varphi_{n,k\alpha} \widetilde  b^\dag_{k\alpha}).
\end{align}
The  amplitudes $\psi_{n}$, $\psi_{n,k\alpha}$ and quasiparticle energies $\Omega_n$ are the solution of the following eigenvalue
problems
\begin{align}\label{sys3}
  &\varepsilon \psi_{n}-t \sum\limits_{k\alpha}\psi_{n,k\alpha}=\Omega_n \psi_{n},
  \notag\\
  &E_{k\alpha}\psi_{n,k\alpha}-  t\psi_{n}=\Omega_n \psi_{n,k\alpha},
\end{align}
whereas
the  amplitudes $\varphi_n$ and $\varphi_{n,k \alpha}$ satisfy the following nonhomogeneous  system of linear equations:
\begin{align}\label{sys4}
   &(\varepsilon-\Omega_n) \varphi_{n}-t \sum_{k\alpha} \varphi_{n,k\alpha}=t \sum_{k\alpha} f_{k\alpha} \psi_{n,k\alpha},
   \notag\\
  &(E^*_{k\alpha}-\Omega_n)\varphi_{n,k\alpha}-  t\varphi_{n}=-t f_{k\alpha} \psi_{n}.
\end{align}
The amplitudes $\psi_n$, $\psi_{n,ka}$ should be normalized according to the first equation in~\eqref{orthogonality}. Operators $\widetilde{c}_n,~\widetilde{c}^{\dagger}_n$ are obtained from~\eqref{a_n2} and~\eqref{a_n22} by the tilde conjugation.
Again, the transformations (\ref{a_n2},\ref{a_n22}) are canonical but nonunitary, so
$\{ c_n, c^{\dagger}_{n'} \}=\delta_{nn'}$ and $ (c_n)^{\dagger} \ne c_n^{\dagger}$, and nontilde and tilde operators anticommute.  Although the analytical expressions for quasiparticle amplitudes and spectrum can be obtained (see Appendix), we found out that it is more convenient for practical calculations and for application to more complex system to solve Eqs.~(\ref{sys3},~\ref{sys4}) numerically.  Namely, we first solve the eigenvalue problem~\eqref{sys3}, and then with the known quasiparticle spectrum $\Omega_n$  and amplitudes  $\psi_{n}$, $\psi_{n,k\alpha}$ we solve the linear system of equations~\eqref{sys4}.

Since the Liouvillian~\eqref{liouv_single} is diagonal in terms  of $c^\dag_n,~c_n$ and their tilde conjugate $\widetilde c^\dag_n,~\widetilde c_n$,
we can associate the vacuum of $c_n,~\widetilde c_n$ with
the  nonequilibrium steady state:
\begin{equation}
c_n |\rho_\infty\rangle =\widetilde{c}_n |\rho_\infty\rangle =0.
\end{equation}
By the construction of $c^\dag_n,~\widetilde c^\dag_n$, $\langle I|c^\dag_n=\langle I|\widetilde c^\dag_n=0.$

Using the transformation inverse to~(\ref{a_n2},~\ref{a_n22}) (see Appendix) we calculate the steady state current and
nonequilibrium electron populations of the molecule and leads levels. For the current, according to~\eqref{current_HF}, we get
\begin{equation}\label{current}
 J_{\alpha} =-2t\,{\rm Im}\sum_{k\,n}\, \psi_{n,k\alpha}\varphi_{n},
\end{equation}
while for the population of the molecule and leads levels we derive
\begin{align}\label{occ}
  &\langle a^\dag a \rangle=\bra{I}a^\dag a\ket{\rho_\infty}=\sum_n \psi_n\varphi_n,
  \notag\\
  &\langle a^\dag_{k\alpha} a_{k\alpha} \rangle =\bra{I}a^\dag_{k\alpha} a_{k\alpha}\ket{\rho_\infty}=f_{k\alpha}+\sum_n \psi_{n,k\alpha}\varphi_{n,k\alpha}.
\end{align}
As is expected, these occupation numbers are  real (see Appendix). Moreover, using the explicit analytical expressions
for amplitudes $\psi_{n,k\alpha},~\varphi_{n,k\alpha}$ it can be demonstrated that the total number of electrons in  both the leads is conserved quantity, i.e.,
 \begin{equation}
   \sum_{k\alpha} \langle a^\dag_{k\alpha} a_{k\alpha} \rangle = \sum_{k\alpha} f_{k\alpha}.
 \end{equation}

In what follows we consider the case of identical left and right leads, i.e.,  $E_{kL}=E_{kR}=E_{k}$. Then
$N$ eigenstates of~\eqref{sys3} coincide with the energies of the leads' levels, i.e.,  $\Omega_n=E_{l}$.
Using analytical expressions for the amplitudes $\psi_{n,k\alpha}$ and $\varphi_n$ (see Appendix) it can be easily shown that only
eigenstates with $\Omega_n=E_l$  contribute the sum over~$n$ in~(\ref{current}). (This  result is  also valid  when $t_{R}=\alpha t_{L}$.)
It is  similar to
Meir and Wingreen observation that the expression for the current can be simplified  within NEGF formalism when the spectral function of the coupling to the
left lead is proportional  to the spectral function of the couplings to the right  lead.\cite{meir92} On the contrary, eigenstates with $\Omega_n=E_l$
do not contribute to the  molecule  population in~\eqref{occ}.

\begin{figure}
 \begin{centering}
\includegraphics[width=15cm]{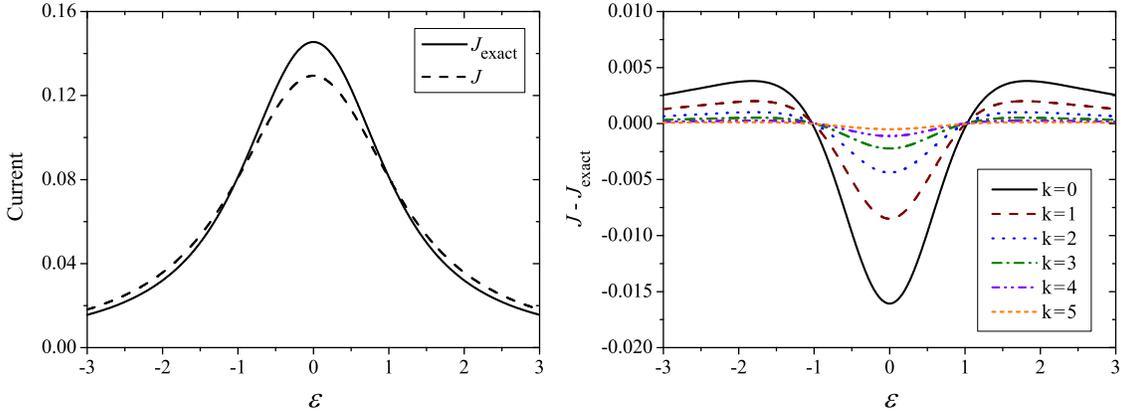}
\caption{(Color) Left panel: The  current  through the one-level molecule as a function of the energy of the level.
 We choose chemical potentials
 of the leads to be $\mu_{L,R}=\pm 0.5$ and $T_{L,R}=0.1$.
 Right panel: The difference between calculated and exact currents for different values
 of $N=100 \times 2^k$.  }
\label{figure1}
 \end{centering}
\end{figure}

\begin{figure}
 \begin{centering}
\includegraphics[width=15cm]{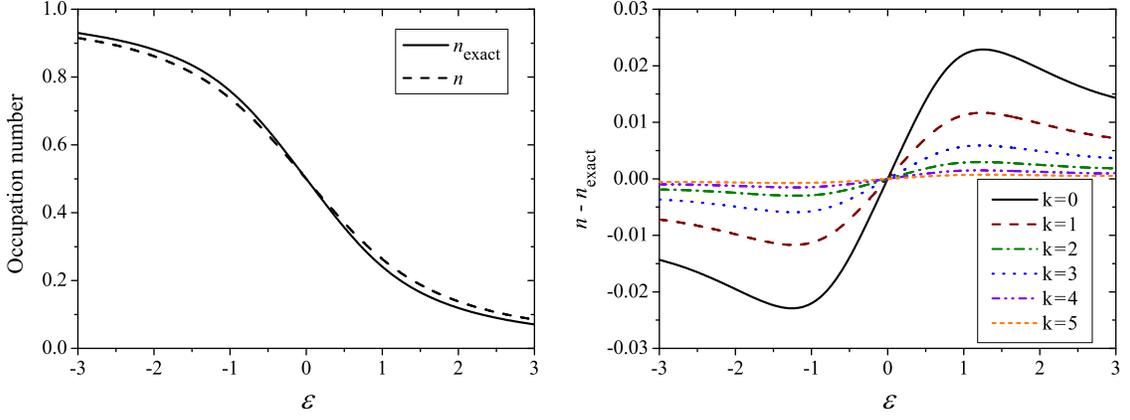}
\caption{(Color)  Left panel: The occupation number [Eq.~\eqref{occ}] for the one-level molecule as a function of  energy of the level.
 Right panel: The difference between  calculated and exact occupation numbers for different values
 of $N$.  The parameters are the same as in Fig.~\ref{figure1}}
 \label{figure2}
 \end{centering}
\end{figure}

Let us compare the results obtained  within our approach with the results given by the Landauer theory.\cite{datta-book}
The Landauer theory is exact for noninteracting electrons and it gives the following expression for the current \cite{datta-book}
\begin{equation}\label{cur_landauer}
  J_{\rm{exact}}=\frac{1}{2\pi} \int d\omega \frac{\Gamma^2(\omega) (f_L(\omega)-f_R(\omega))}{(\omega-\varepsilon-2\Lambda (\omega))^2 +(\Gamma(\omega))^2},
\end{equation}
and for the population
\begin{equation}
\label{exact-n}
  n_{\rm{exact}}=\frac{1}{2\pi} \int d\omega \frac{\Gamma(\omega) (f_L(\omega)+f_R(\omega))}{(\omega-\varepsilon-2\Lambda (\omega))^2 +(\Gamma(\omega))^2}.
\end{equation}
Here  $\Lambda (\omega)$ and $\Gamma(\omega)$ are real and imaginary parts of the self-energy of the leads
\begin{equation}\label{SelfEnergy}
\Sigma(\omega)=  t^2 \sum_{k }\frac{1}{\omega-\varepsilon_{k\alpha}+i\gamma}= \Lambda (\omega) - \frac{i}{2}\Gamma(\omega).
 \end{equation}
Notice that the self-energy~\eqref{SelfEnergy} exactly appears in secular  equation \eqref{secular-t}.
We do not distinguish between the self-energies of the left and right leads, since
the both leads  are identical. In numerical calculations by the Landauer theory we put $N\to\infty$,  $\gamma\to 0$ and assume
that leads densities of states are constant within the energy bandwidth $[E_{min}:E_{max}]$.
Then $\Gamma(\omega)=\Gamma$ for $E_{min}\leq\omega\leq E_{max}$ and zero otherwise. For $\Lambda(\omega)$ we have the Cauchy principal value integral
\begin{equation}
  \Lambda(\omega)=\frac{1}{2\pi} \; \text{P} \int\limits_{-\infty}^{\infty}\frac{\Gamma(\varepsilon)d\varepsilon}{\omega-\varepsilon}=
    \frac{\Gamma}{2\pi}\ln\left|\frac{\omega-E_{min}}{\omega-E_{max}}\right|.
\end{equation}

In our calculations with the kinetic equation we assume that $N$ energy levels in each lead are evenly spaced in the bandwidth $[E_{min}:E_{max}] =[-5:5]$.  The tunneling coupling strength $t$  is computed from the
$\Gamma=2\pi\eta t^2=1$, where $\eta=N/(E_{max}-E_{min})$ is the density of states.
 Below we put $\gamma=2\Delta\varepsilon$, where $\Delta\varepsilon$
is the energy spacing between states in the leads. Under this choice of the parameters $\Gamma(\omega)$,  as is computed by (\ref{SelfEnergy}), is  constant equal to $1$  within the bandwidth of the leads and vanishes when $\omega$ is outside the bandwidth.

In the left panel of Fig.~\ref{figure1}, we plot the current~[Eq.~\eqref{current}] calculated within our approach for $N=100$ as a function of the level energy along with the exact current computed by the Landauer formula~\eqref{cur_landauer}.
We see that the calculated current agrees well with the exact one. The largest
deviation between the two currents is obtained when the currents reach their maxima. In the right panel of Fig.~\ref{figure1} we plot the difference between the calculated within our approach
and exact currents for  different values of $N$. It is evident from the figure that the difference becomes smaller as the leads densities of states increase.
It agrees with our observation (see Appendix)  that our expression for the current~\eqref{current} becomes the Landauer formula~\eqref{cur_landauer} in the limit of macroscopically large leads.
We also calculate  nonequilibrium population of the molecule and compare them with the exact values (\ref{exact-n}) for  different values of $N$. The results are shown in Fig.~\ref{figure2}. Again, we reproduce the exact results by increasing the part of the leads included into the consideration.

\subsection{Out of equilibrium Anderson model in Hartree-Fock approximation}

In this section, we apply our method to a physically more interesting example, namely to electron transport  through a  spin-degenerate single level  with local Coulomb interaction (so called Anderson model).  The  Hamiltonian for the  molecule has the form
\begin{equation}
  H_S= \varepsilon \sum_{\sigma} a^\dag_\sigma a_\sigma  + U a^\dag_\uparrow a_\uparrow a^\dag_\downarrow a_\downarrow,
\end{equation}
where $a^\dag_\sigma,~a_\sigma$ are the creation and annihilation operators for the spin-up ($\sigma=\uparrow$) and spin-down ($\sigma=\downarrow$) electrons in the molecule,
$\varepsilon$ is the energy of the single level in the molecule, and the charging energy $U$ characterizes
the Coulomb interaction of electrons in the molecule. Here we restrict our consideration by a nonmagnetic system, therefore  the energy levels in the leads are spin degenerates and  all the coupling strengths $t_{k\alpha\sigma}$ take the same value $t$. Under these assumptions we have
\begin{equation}
H_B + H_T= \sum_{k \alpha \sigma} \varepsilon_{k\alpha}a^{\dagger}_{k \alpha\sigma}a_{k \alpha\sigma} -
 t\sum_{k \alpha \sigma} (a^{\dagger}_{k \alpha\sigma} a_\sigma + h.c.).
\end{equation}
We are interested in quantum transport in Coulomb blockade regime, where we charging energy $U$ is much greater than the effective coupling
$\Gamma_{\uparrow,\downarrow}=2\pi\eta t^2$ between the molecule and the leads.  So we work in the regime of strong Coulomb interaction and weak coupling to the leads.

Using the Wick theorem with respect to nonequilibrium steady state vacuum  $\ket{\rho^{(0)}_\infty}$ (which is yet to be defined) from the right and the left vacuum $\langle I | $, we obtain the  Hartree-Fock  part of $H_S$ in the following form
\begin{equation}
  H^0_S = \sum_\sigma (\varepsilon +U n_{-\sigma})  a^\dag_\sigma a_\sigma - U( n_{\downarrow\uparrow} a^\dag_\downarrow a_\uparrow + n_{\uparrow\downarrow}a^\dag_\uparrow a_\downarrow)
\end{equation}
where $n_{\sigma}=\bra{I} a^\dag_{\sigma}a_\sigma\ket{\rho^{(0)}_\infty}$ and $n_{\sigma\sigma'}=\bra{I} a^\dag_{\sigma'}a_\sigma\ket{\rho^{(0)}_\infty}$. Since the system is nonmagnetic, the occupation numbers for both spin orientations are the same
$n_{\sigma}=n_{-\sigma}$. Moreover, due to the symmetry reasons  $n_{\uparrow\downarrow}=n_{\downarrow\uparrow}$. Then the following unitary  transformation
\begin{equation}
\alpha_1^\dag=\frac{1}{\sqrt2}(  a^\dag_\uparrow- a^\dag_\downarrow) ,~~~ \alpha_2^\dag=\frac{1}{\sqrt2}(  a^\dag_\uparrow + a^\dag_\downarrow)
\end{equation}
brings $H^0_S$ to a diagonal form
\begin{equation}
  H^0_S=\sum_{s=1,2} \varepsilon_ s \alpha^\dag_s\alpha_s,
\end{equation}
where the Hartree-Fock mean-field energies are
\begin{align}\label{coul_HF}
  \varepsilon_{1(2)}=\varepsilon + U(n_{\sigma}\pm n_{\uparrow\downarrow})=\varepsilon + U n_{2(1)},
\end{align}
and $n_{s}=\bra{I} \alpha^\dag_s \alpha_s\ket{\rho^{(0)}_\infty}$ is the occupation number for the Hartree-Fock single-particle level "$s$" (the crossover matrix element
$\bra{I} \alpha^\dag_{s} \alpha_{s'\ne s}\ket{\rho^{(0)}_\infty} $ is equal zero) and $s=1(2)$ corresponds to $"+(-)"$ signs respectively. We can regard $\alpha_1^{\dagger}$ as the antibonding orbital creation operator,  and
 $\alpha_2^{\dagger}$ as the creation operator for the bonding orbital in the dot.
 From the above equation it follows that if $n_{\uparrow\downarrow}\ne0$,  the Hartree-Fock bonding/antibonding energies $\varepsilon_{1(2)}$ are split and the populations of bonding/antibonding orbitals are generally not the same $n_1\ne n_2$.

\begin{figure}
 \begin{centering}
\includegraphics[width=15cm]{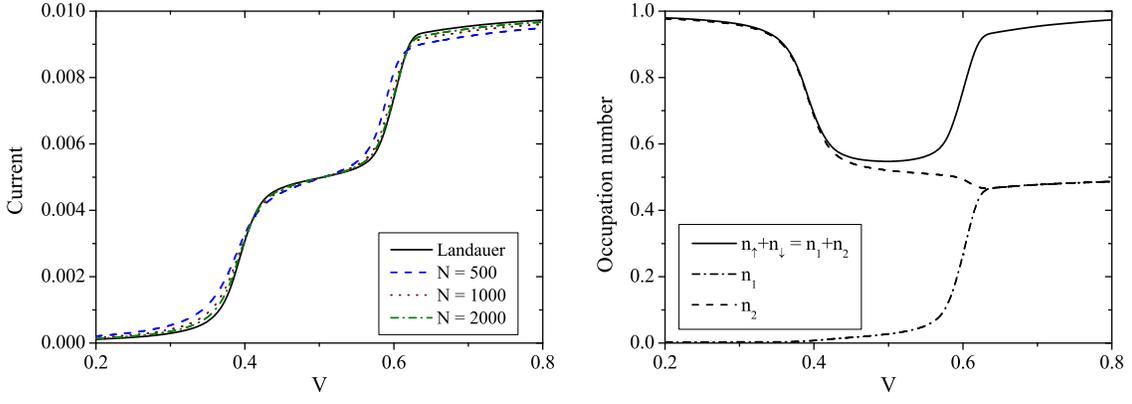}
\caption{(Color) The current (left panel) and the occupation numbers (right panel) for out of equilibrium Anderson model at fixed
gate voltage $V_g=-0.2$ as a function of voltage difference between left and right leads. For the current we show the results obtained with  different number of states $N$ in each lead and the exact Hartree-Fock result.
In the right panel the total occupation number and  the nonequilibrium populations of the Hartree-Fock levels  are shown for $N=2000$. }
 \label{figure3}
 \end{centering}
\end{figure}

\begin{figure}
 \begin{centering}
\includegraphics[width=15cm]{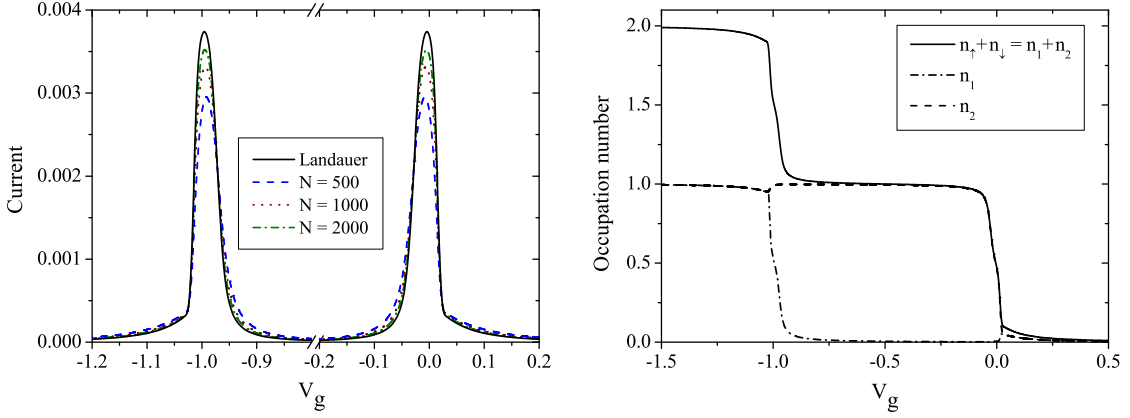}
\caption{(Color) The current (left panel) and the occupation number (right panel) for out of equilibrium Anderson model at small applied voltage
$\mu_{L,R}=\pm 0.025$ as a function of gate voltage.  The parameters and notations are the same as in Fig.~\ref{figure3}.}
 \label{figure4}
 \end{centering}
\end{figure}

Following the procedure described in Sect.~\ref{section3} we introduce  $b^\dag_s$ and $b^\dag_{k\alpha\sigma}$ operators which annihilate the left vacuum $\bra{I}$.
Then the quadratic part of the Liouvillian becomes
\begin{align}\label{liouv_Coulomb}
L^{(0)}= &\sum_s \varepsilon_s(b_s^\dag b_s-\widetilde b_s^\dag \widetilde b_s) +  L_B
\notag \\
&-\sum_{k\alpha\sigma s} T_{s\sigma }\left\{(b^\dag_{k\alpha\sigma}b_s+b^\dag_s b_{k\alpha\sigma})-
 (\widetilde b^\dag_{k\alpha\sigma}\widetilde b_s+\widetilde b_s^\dag \widetilde b_{k\alpha\sigma})+
 if_{k\alpha}(b^\dag_{k\alpha\sigma}\widetilde b^\dag_s+\widetilde b_{k\alpha\sigma}^\dag b^\dag_s ) \right\},
\end{align}
where $T_{1\downarrow}=-t/\sqrt{2}$ and all other $T_{s\sigma}$ are equal $t/\sqrt{2}$. Likewise the noninteracting single-level  model considered in the previous section, the  Liouvillian~\eqref{liouv_Coulomb}
can be diagonalized in terms of nonequilibrium quasiparticles by the equation of motion method.  The structure of nonequilibrium quasiparticles, which diagonalize~\eqref{liouv_Coulomb} is the following ($n=1,\ldots, 2+4N$)
\begin{align}\label{c_Coul}
  c^\dag_n&=\sum_{s}\psi_{n,\,s}b^{\dagger}_s+\sum_{k\alpha\sigma}\psi_{n,\,k\alpha\sigma}  b^\dag_{k\alpha\sigma},~~\widetilde c^\dag_n=(c^\dag_n)\widetilde{},
   \notag \\
  c_n&=\sum_{s}(\psi_{n,\,s}b_s +  i   \varphi_{n,\,s} \widetilde b^\dag_s)+\sum_{k\alpha\sigma}(\psi_{n,k\alpha\sigma}  b_{k\alpha\sigma} +i \varphi_{n,k\alpha\sigma} \widetilde  b^\dag_{k\alpha\sigma}),~~~\widetilde c_n=(c_n)\widetilde{},
\end{align}
where the amplitudes $\psi$, $\varphi$ and the spectrum $\Omega_n$ are obey of the equations ($s=1,~2$)
\begin{align}\label{sys3Coul}
  &\varepsilon_s \psi_{n,s}- \sum\limits_{k\alpha\sigma}T_{s\sigma}\psi_{n,k\alpha\sigma}=\Omega_n \psi_{n,s},
  \notag\\
  &E_{k\alpha}\psi_{n,k\alpha\sigma}-  \sum_{s}T_{s\sigma}\psi_{n,s}=\Omega_n \psi_{n,k\alpha\sigma},
\end{align}
and
\begin{align}\label{sys4Coul}
   &(\varepsilon_s-\Omega_n) \varphi_{n,s}- \sum_{k\alpha\sigma} T_{s\sigma} \varphi_{n,k\alpha\sigma}= \sum_{k\alpha} t_{s\sigma}f_{k\alpha} \psi_{n,k\alpha},
   \notag\\
  &(E^*_{k\alpha}-\Omega_n)\varphi_{n,k\alpha\sigma}- \sum_s T_{s\sigma}\varphi_{n,s}=-f_{k\alpha} \sum_s t_{s\sigma}\psi_{n,s}.
\end{align}
 The amplitudes depend on Hartree-Fock occupations numbers $n_s$, which in turn depend on the
amplitudes:
\begin{equation}
  n_s=\sum_n  \psi_{n,\,s}\varphi_{n,\,s}.
\end{equation}
Therefore the  equations (\ref{sys3Coul},\ref{sys4Coul})  should be solved via self-consistent iterations. After the diagonalization, the  Liouvillian~\eqref{liouv_Coulomb} takes the form~\eqref{L_op2}. Having found  the amplitudes for the nonequilibrium quasiparticles we can compute the nonequilibrium population of the molecule  ($n_\uparrow+n_\downarrow=n_1+n_2$) as well as the steady state current~\eqref{current_HF}.

Likewise the example with single molecule level, in case of identical leads  the system~\eqref{sys3Coul} have $N$ twice degenerate eigenstates
equal to the energy levels in the leads. Only these eigenstates contribute to the  steady state current,  but they do not contribute to the
population of the molecule.

In our further numerical calculations we assume the symmetrical
voltage drop, $\mu_{L,R}=E_F\pm\frac12V$, and $E_F=0$.  Also, we introduce gate voltage $V_g$, so that the energy of molecular level  is $\varepsilon(V_g)=E_F+V_g$.
We put $U=1.0$ for the strength of the Coulomb interaction. The leads energy levels are spaced in the bandwidth [-1:1] and $\Gamma_{\uparrow,\downarrow}=0.01$,
$T_{L,R}=0.001$.

Fig.~~\ref{figure3} shows the calculated current through the molecule and the occupation numbers as function of the bias voltage applied across the junction  at fixed gate voltage $V_g=-0.2$.  While both chemical potentials are above the molecular energy level $\varepsilon=-0.2$  the electric current through the molecule is small and the average occupation number of the molecule is close to unity. It means that the bonding Hartree-Fock level $s=2$ is below  $\mu_{L,R}$ ($\varepsilon_2=\varepsilon$)   and it is totally occupied ($n_2=1$) while the antibonding level $s=1$ is above $\mu_{L,R}$  ($\varepsilon_1=\varepsilon+U$) and it is empty ($n_1=0$).
 The first step in the current occurs  when the right chemical potential reaches the energy $\varepsilon$ ($V\approx0.4$) and the Hartree-Fock bonding level $s=2$   becomes involved in electronic transport. As a result the occupation of this level decreases to $0.5$, and the energy of the empty Hartree-Fock level $s=1$  reduces  to $\varepsilon_1=\varepsilon + 0.5 U$. The total number of electrons is the molecule becomes $n_1+n_2=0.5$.
The second step in the current occurs when the left chemical potential reaches the  energy level $\varepsilon_1=\varepsilon + 0.5 U$ ($V\approx0.6$)  for the antibonding orbital and electrons starts to flow through it. This increases the occupation number of the antibonding Hartree-Fock orbital level $s=1$ from zero to $0.5$ and, therefore, the bonding level $s=2$ is pushed up from $\varepsilon_2=\varepsilon$ to  $\varepsilon_2=\varepsilon + 0.5 U$.
At higher voltage ($V>0.6$)  the  Hartree-Fock levels become degenerate and they both lay between $\mu_L$ and $\mu_R$. Since each of them is half-occupied, the total number of electrons in the molecule becomes again equal to unity.

Fig.~~\ref{figure4} shows the current and the occupation numbers as a function of the gate voltage $V_g$, at small applied voltage $\mu_{L,R}=\pm 0.025$.
At $V_g<-1.0$ both Hartree-Fock levels are  below ($\varepsilon_{1,2}=\varepsilon+U<0$)
 the chemical potentials  for both leads. Accordingly, the occupation number is equal to two and the current through the junction is close to zero.
At $V_g\approx-1.0$ the molecule starts to conduct electrons This result in a pronounced increase in the current accompanied by a  decrease in the occupation number.
At $-1.0<V_g<0$ the Hartree-Fock bonding level is below chemical potentials ($\varepsilon_2=\varepsilon$), and antibonding level is above them ($\varepsilon_1=\varepsilon+U$).
Therefore the occupation number is equal to unity and the current through the molecule  becomes almost zero At $V_g\approx0.0$ the bonding level becomes involved in the electronic transport giving rise to the current increase. At higher gate voltage, when the single-particle level $\varepsilon$ is above the chemical potentials, the molecule become empty and the current again drops to zero.

In~Figs.~\ref{figure3} and~\ref{figure4} we  compare our results with the exact Landauer current through Hartree-Fock levels. To obtain the exact current we first
solve iteratively the nonlinear equations $n_{1(2)}=n_{exact}(\varepsilon_{1(2)})$ (\ref{exact-n}). Then the Hartree-Fock levels $\varepsilon_{1(2)}$ are used in~\eqref{cur_landauer} to
compute the current. Figs.~\ref{figure3} and~\ref{figure4} show that  the currents calculated within  the present approach converge to the exact result with increasing value of $N$.

\section{Conclusions}

Based on the super-fermion representation of the Liouville space, we developed an approach which enables us to transform the
quantum master equations to the super-Fock space and then to use  standard methods of quantum field theory to solve them in the nonequilibrium steady state regime.
We worked with the Lindblad master equation in this paper although the derivations can be readily extended to more sophisticated master equations.
The main technical difficulty of the approach is that the left vacuum is always different from the right vacuum state unless we have a system in thermodynamic equilibrium.
This prohibits the use of the unitary transformations to diagonalize the Liouvillian. The problem was circumvented  by the development of  a set of nonunitary, canonical  transformations between particle creation and annihilation operators.  These nonunitary canonical transformations preserve the anticommutation relations between the fermionic operators  and  significantly facilitate  the derivations.

Starting with Lindblad master equation for the electron transport through the interacting region, we converted the problem of finding the nonequilibrium steady state  to the many-body problem with non-Hermitian "Hamiltonian" in super-Fock space. Then we demonstrated that despite the fact that the left vacuum is different from the right vacuum we still can use the Wick theorem. Using the Wick theorem we transformed the Liouvillian to the normal ordered form, introduced nonequilibrium quasiparticles and developed a general many-body theory for  electron transport through  interacting region. We applied the approach to  electron transport through a single level molecule. Then we considered the system with electron-electron interactions, namely out of equilibrium Anderson model in Hartree-Fock approximation.  The Wick theorem was applied to obtain the Hartree-Fock solution of the transport problem. We demonstrated that it is consistent with the  Landauer theory, which is exact  for these models.
Being formulated in the language of  Fock spaces, creation and annihilation operators and  normal ordered "Hamiltonian" the proposed approach is not only capable of doing perturbative calculations but also  has a great complementarity to the nonperturbative many-body methods of molecular electronic structure calculations such as, for example,  coupled cluster or configuration interaction theories.

\begin{acknowledgments}
The authors thank M. Esposito, M. Gelin,  and T. Prosen for  valuable discussions. This work has been supported by the Francqui Foundation and by the Belgian Federal Government under the
Inter-university Attraction Pole project NOSY P06/02.
\end{acknowledgments}

\newpage
\appendix
\section{Structure of nonequilibrium quasiparticles creation and annihilation operators for single level model}

For the single level model the analytical solution of Eqs.~(\ref{sys3},~\ref{sys4})  has the following form:
\begin{equation}\label{psi_single2}
\psi_{n}={\cal N}_n^{-1/2},~~\psi_{n,k\alpha}=\frac{t}{E_{k\alpha}-\Omega_n}\psi_{n},
\end{equation}
\begin{align}\label{varphi_single2}
\varphi_{n}={\cal N}_n^{-1/2}
                    \cfrac{\sum\limits_{k\alpha}\cfrac{f_{k\alpha}\gamma_{k\alpha}}{(E_{k\alpha}-\Omega_n)(E^*_{k\alpha}-\Omega_n)}}
         {\sum\limits_{k\alpha}\cfrac{\gamma_{k\alpha}}{(E_{k\alpha}-\Omega_n)(E^*_{k\alpha}-\Omega_n)}}
           ~,~~~~~~
  \varphi_{n,k\alpha}=\frac{t}{E^*_{k\alpha}-\Omega_n}\bigl(\varphi_{n}- \psi_n f_{k\alpha} \bigr)~.
\end{align}
where
${\cal N}_n=1+t^2\sum\limits_{k\alpha}(E_{k\alpha}-\Omega_n)^{-2}$, and  $\Omega_n$ obey the secular equation
\begin{equation}\label{secular-t}
  (\varepsilon-\Omega)+t^2\sum_{k\alpha}\frac{1}{\Omega-E_{k\alpha}}=0.
\end{equation}

Some constraints on quasiparticle amplitudes can be obtained by using the transformations inverse to~(\ref{a_n2},\ref{a_n22}):
\begin{align}
\label{inv-transf}
&  b^\dag =\sum_n\psi_{n }c^\dag_n,~~ b=\sum_n(\psi_{n}c_n-i\varphi^*_{n}\widetilde c^\dag_n),
   \notag\\
&  b^\dag_{k\alpha}=\sum_n \psi_{n, k\alpha }c^\dag_n,~~ b_{ k\alpha }=\sum_n(\psi_{n, k\alpha }c_n-i\varphi^*_{n, k\alpha}\widetilde c^\dag_n)
\end{align}
(the expressions for $\widetilde{b},~\widetilde{b}^{\dagger},~ \widetilde{b}_{k \alpha},~\widetilde{b}^{\dagger}_{k\alpha}$ are obtained from (\ref{inv-transf}) by the tilde conjugation). In particular, from $\{b_{k\alpha},\widetilde b_{k\alpha}\}=0$ and $\{b,\widetilde b\}=0$ it follows that
$\sum\limits_n \psi_{n,k\alpha}\varphi_{n,k\alpha}$ and $\sum\limits_n \psi_n\varphi_n$ are real.

If we have identical left and right leads, i.e., if $E_{kL}=E_{kR}=E_{k}$, then
$N$ eigenstates of~\eqref{sys3} coincide with the energies of the leads levels, i.e.,  $\Omega_n=E_{l}$. For such normalized eigenstates we have
\begin{equation}\label{psi_single1}
  \psi_n=0,~~~~\psi_{n,kL}=- \psi_{n,kR}=\delta_{kl}\frac{1}{\sqrt 2},
 \end{equation}
where $\delta_{kl}$ is the Kroneker symbol, and
\begin{align}\label{phi_single}
 &\varphi_{n}= 2^{-1/2}
                  \cfrac{t(f_{lL}-f_{lR})}
         {(\varepsilon-E_l)+2 t^2\sum\limits_{k}\cfrac{1}{(E_l-E^*_{k})}}~,~~~~\varphi_{n,k\alpha}=\cfrac{t\varphi_{n}}{(E^*_{k}-E_l)}.
                     \end{align}
This  result is  also valid  when $t_{R}=\alpha t_{L}$.
The only difference is that $\psi_{n,\,lR}= -\alpha \psi_{n,\,lL}$. Therefore, in case of identical leads the summation over $n$ in~\eqref{current}
can be divided into two parts. The first part involves summation over $N+1$ solutions of Eq.~\eqref{secular-t}, while the second part involves summation over $N$ solutions
 of~\eqref{sys3} such that $\Omega_n=E_{l}$. The first  part does not depend on index $\alpha$ and it vanishes. Therefore
\begin{equation}\label{current_mol}
 J_{\alpha} =\mp{\rm Im}\sum_{l}
                  \cfrac{t^2(f_{lL}-f_{lR})}
         {(\varepsilon-E_l)+2 t^2\sum\limits_{k}\cfrac{1}{(E_l-E^*_{k})}}~.
\end{equation}
Here the upper (lower) sign corresponds to $\alpha=L$ ($\alpha=R$). It is obvious that in the limit of macroscopically large  leads ($N\to\infty,~\gamma\to0$)
eq.(\ref{current_mol})  becomes the standard Landauer formula~\eqref{cur_landauer} for the current.

\newpage
\bibliographystyle{apsrev}

\end{document}